## *Ultra-compact and Wide-spectrum-range Thermo-optic Switch Based on Silicon Coupled Photonic Crystal Microcavities*

Xingyu Zhang,[1,a)] Swapnajit Chakravarty,[2,a)] Chi-Jui Chung,[1] Zeyu Pan,[1] Hai Yan,[1] and Ray T. Chen[1,2,a)]

[1]*Department of Electrical and Computer Engineering, University of Texas at Austin, Austin, TX, 78758, USA*

[2]*Omega Optics, Inc., Austin, TX 78757, USA*

We design, fabricate and experimentally demonstrate a compact thermo-optic gate switch comprising a 3.78 μm-long coupled L0-type photonic crystal microcavities on a silicon-on-insulator substrate. A nanohole is inserted in the center of each individual L0 photonic crystal microcavity. Coupling between identical microcavities gives rise to bonding and anti-bonding states of the coupled photonic molecules. The coupled photonic crystal microcavities are numerically simulated and experimentally verified with a 6nm-wide flat-bottom resonance in its transmission spectrum, which enables wider operational spectrum range than microring resonators. An integrated micro-heater is in direct contact with the silicon core to efficiently drive the device. The thermo-optic switch is measured with an optical extinction ratio of 20dB, an on-off switching power of 18.2mW, a therm-optic tuning efficiency of 0.63nm/mW, a rise time of 14.8 μsec and a fall time of 18.5 μsec. The measured on-chip loss on the transmission band is as low as 1dB.

Integrated optical switches are important building blocks in silicon photonics. While output-port-selective switches are useful for optical routing applications, optical gate switches with single output port for off-to-on and on-to-off operations also have many potential applications such as optical interconnects and optical logic devices [1,2]. Silicon based optical gate switches have attracted a significant amount of attentions in recent years, due to their small size, large scalability, and potential for integration with wavelength-division-multiplexing (WDM) systems. Since silicon's thermo-optic (TO) effect is significantly larger than its electro-optic (EO) effect, silicon TO switches promises efficient and low-power operation. Conventional Mach–Zehnder interferometers and directional couplers can be used as optical gate switching structures, but they require long waveguides (several millimeters) to obtain π phase shift of light and require relatively high switching power, which limits the integration density and energy conservation [3,4]. Another type of widely-used compact structure for efficient switches a microring resonator. So far, the smallest ring diameter reported is 3 μm [5]; however, one drawback of microring switches is the very narrow operational optical bandwidth (<1nm), as well as the requirements of resonance wavelength stabilization techniques [6]. To address this issue, cascaded microrings along one waveguide have been developed to broaden the overall optical bandwidth [7]; however, this not only increases the total device footprint and system complexity, but also increases the total power consumption. Furthermore, although all these microrings are designed to be ideal, the randomness in fabrication may make them slightly off-resonance among each other and with different extinction ratios; therefore, resonance tuning on each individual microring is required [8]. In recent years, some photonic crystal (PC)-based devices on silicon-on-insulator (SOI) substrates have been proposed as compact and low-power switches [9,10]; however, the dispersion diagram needs to be carefully engineered to maximize the extinction ratio, and its operational optical bandwidth is limited by the high dispersion of PCs. Another type of switch is based on a band-edge shifted photonic crystal waveguide (PCW) [11,12]; nevertheless, it needs to be operated very close to the PCW transmission band edge where the large group index leads to high optical loss, and large PCW dispersion near the band edge limits the operational optical bandwidth. The band edge shifted PCW also requires larger heating power and temperature shifts for switching. Other switch structures based on waveguide Fabry-Perot microcavities [13] and a single PC nanocavity [14] suffer from narrow operational optical bandwidth due to inherent limitation from their narrow resonance linewidths. Therefore, a TO switch with compact size, wide optical bandwidth, low loss and low switching power, as well as ease of design and fabrication, is highly required.

---

[a] Authors to whom correspondence should be addressed. Email addresses: xzhang@utexas.edu, swapnajit.chakravarty@omegaoptics.com and chenrt@austin.utexas.edu

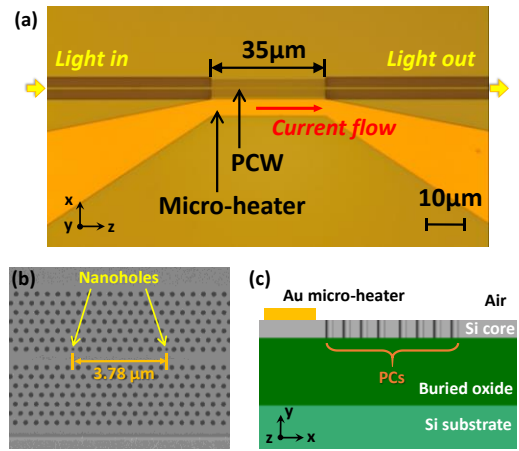

FIG. 1. (a) Fabricated thermo-optic switch, consisting of a W1 PCW with coupled PC microcavities, integrated with a micro-heater, on an SOI substrate. (b) SEM image of the W1 PCW and 3.78 µm-long PC microcavities inserted with nano-holes. (c) The schematic of the cross section of the PCW and the micro-heater.

In this paper, we design, fabricate and characterize a compact and high-performance TO switch based on two coupled L0-type PC microcavities integrated with a micro-heater on a SOI substrate. Various coupled PC microcavities have been theoretically studied and experimentally demonstrated previously for switching in the context of gap solitons [15] and optical analogs to electromagnetically induced transparency [16]. However, the coupled cavity geometry has rarely been explored as an optical switch. In this work, simulations are performed to design coupled PC microcavities with a 6nm-wide resonance dip in the transmission spectrum. This structure is theoretically analyzed in terms of the bonding and anti-bonding states of homoatomic photonic molecules. Measurement results on the fabricated coupled PC microcavities verify the 6nm-wide resonance dip and also shows an optical extinction ratio of ~20dB. For data transmission or gate switching, a flat-top and a flat-bottom in the transmission spectrum are preferred in order to minimize the signal waveform distortion and thus the data transmission errors. For DWDM add-drop functions, flat top and flat bottom are preferred since they provide a much better 1-dB and 3-dB passband [17]. Therefore, microring resonators are not favorable due to their sharp Lorentzian line shape with narrow optical bandwidth and non-flat-bottom [18]. Instead, our PC microcavities with flat-bottom resonance can address this issue, providing significantly broader operational optical bandwidth. Furthermore, a wide flat-bottom resonance is achieved with only two coupled PC resonators compared to widely-used multi-cascaded microring resonators, thereby enabling significantly small footprints. A micro-heater is designed in this work to efficiently drive the device, and its heater transfer characteristics is numerically simulated. The resonance shifts are measured under different DC driving power, and a constant switching power for this 6nm-wide operational spectrum is measured to be 18.2mW. The TO tuning efficiency is measured as 0.63nm/mW. On-off switching is performed, and a rise time of 14.8 μsec and a fall time of 18.5 μsec are measured. The on-chip optical loss is measured to be about 1dB. While TO tuning is used here to demonstrate the functional switching on the coupled PC microcavities, higher switching speed could be achieved by free carrier modulation in the future, enabling more potential applications including high-speed photonic interconnects and broadband optical communication systems.

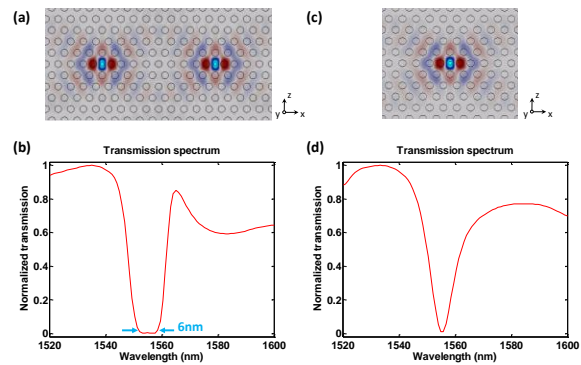

FIG. 2. (a) Simulated out-of-plane magnetic field ($H_y$) component of a transverse electric (TE) confined resonant mode in an isolated coupled L0-type PC microcavities. The field profile is for the anti-bonding mode resonance. (b) Simulated transmission spectrum of the coupled PC microcavity in (a) adjacent to a W1 PCW, showing a 6nm-wide resonance dip in the transmission band. (c) Simulated out-of-plane magnetic field ($H_y$) component of a TE confined resonant mode in an isolated L0-type PC microcavity. (d) Simulated transmission spectrum of the PC microcavity coupled to a W1 PCW, as a comparison.

Figs. 1 (a) and (b) show our fabricated TO switch on an SOI substrate [Si thickness=250nm, buried oxide thickness=3μm, as illustrated in Fig. 1 (c)]. A 35 μm-long W1 PCW is designed with a lattice constant of $a$=420nm, hole diameter of $d$=201nm, and with 7 rows of holes on each side. The input and output strip waveguides are connected to the PCW using low-loss group index tapers, in order to reduce the optical loss caused by group index mismatch [19]. Grating couplers are used to couple light into and out of the chip. As shown in Fig. 1 (b), two L0-type PC microcavities are formed by shifting two nearby air holes adjacent to the W1 PCW by 0.2$a$=84nm from their lattice positions away from each other. These two PC microcavities are separated by 9 periods of holes along the direction parallel to the W1 PCW. Each microcavity is further modified by inserting a nanohole [20] with diameter of 0.5$d$=100.5nm. The length of our coupled PC microcavities is only 3.78 μm. Figure 2 (a) shows the out-of-plane magnetic field ($H_y$) component of a transverse electric (TE) confined resonant mode in our isolated coupled PC microcavity. When the coupled PC microcavities are located along the Γ−K direction parallel to the W1 PCW [Fig. 1 (b)], the coupling of the two resonators leads to a 6nm-wide resonance dip in the transmission spectrum, with a nearly flat valley, as shown in the 3D FDTD (finite difference time domain) simulated transmission spectrum in Fig. 2 (b). The flat-bottom resonance in Fig. 2 (b) is located ~70nm away from the W1 PCW band edge (at ~1620nm). The group index in this low-dispersion region is ~6, which is much lower compared to band-edge shifting switches [11]. The maximum transmission on the left edge of the resonance also shows an almost smooth top (from 1532nm to 1538nm), which is important for signal distortion minimization in TO switching process. As a comparison, Fig. 2 (c) shows the out-of-plane magnetic field ($H_y$) component of a TE confined resonant mode in an isolated L0-type PC microcavity formed by shifting two adjacent air holes by 0.2a and inserting a nanohole with a diameter of 0.5d at the center of each cavity. The resonance mode has a quality factor ~6,924 and volume of ~0.26 $(\lambda/n)^3$. Figure 2 (d) shows the drop transmission spectrum when the PC microcavity is coupled to a W1 PCW. In both the coupled and the uncoupled L0-type PC microcavities, it is observed that the field maximum overlaps completely with the central nanohole.

Coupled PC microcavities are also called photonic molecules. Coupling between identical localized photonic modes in our identical L0-type PC microcavities or L0-type homoatomic molecules give rise to a frequency splitting into a bonding state and an anti-bonding state [21]. Figures 3 (a)-(c) show the simulated magnetic field component for bonding and anti-bonding states of the split resonances in our isolated coupled L0-type PC microcavities, separated by 2, 3 and 4 periods, respectively. At a separation of 3 periods, the resonant modes in the bonding and anti-bonding states have quality factors of ~9,660 and ~7,484, respectively. At separation greater than 4 periods, the bonding and anti-bonding states are almost degenerate. In quantum mechanics, the molecular ground state is expected to be a bonding state and the first excited state is expected to be an anti-bonding state. However, due to the oscillating nature of the evanescent waves in the photonic band gaps, as observed in other photonic systems [22,23], and also in electronic systems [24], in our system of coupled L0-type PC microcavities aligned at 0 degrees, the lower frequency ground state changes from bonding to anti-bonding by varying the distance between the PC microcavities, as shown in Fig. 3 (e). When the photonic molecules are coupled to a W1 PCW, the transmission drop spectrum is a function of both the resonance spacing and the dispersion of the guided mode of the W1 PCW. The evolution of the simulated transmission spectrum as a function of the number of PC hole periods (P) between the coupled microcavities is analyzed, as shown in Fig. s1 [25]. Based on these simulation results, coupled L0-type PC microcavities with a separation of 9 lattice periods (P=9) was chosen in the device design in this work, since this structure provides the widest resonance dip with the shortest cavity length. Figure 3 (d) shows the coupled resonant modes at the separation of 9 PC lattice holes, for bonding and anti-bonding states. Furthermore, we also investigated the variation of the transmission spectrum as a function of the PC hole diameter, and simulation results in Fig. s2 show that diameter variations from -10nm to +10nm lead to resonance shifts within +18nm to -18nm, while the width of resonance dip remains almost unchanged [26].

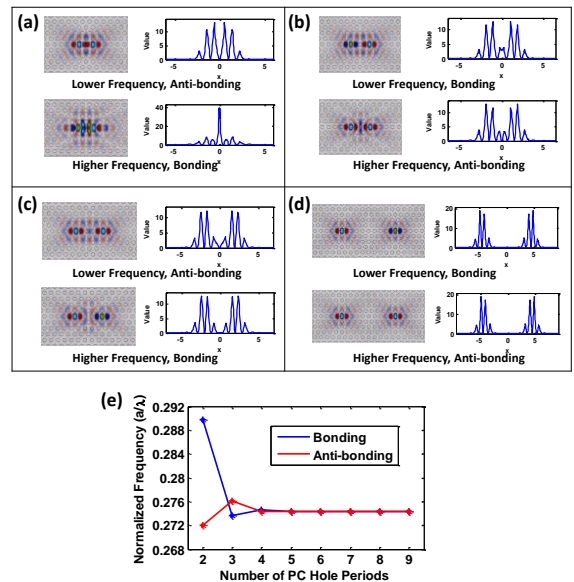

FIG. 3. Simulated out-of-plane magnetic field component for TE confined bonding and anti-bonding resonance states of the two coupled L0-type PC microcavities separated by (a) 2, (b) 3 (c) 4, and (d) 9 periods, respectively. The blue curves in (a)-(d) show the in-plane confined transverse electric field intensity of the modes at the cross-section along the line passing through the two microcavities. (e) Simulated frequencies of the bonding and anti-bonding states as a function of the separation between the coupled L0-type PC microcavities.

To efficiently drive the device through TO effect, a gold strip micro-heater electrode [26] is designed, as shown in Figs. 1 (a) and (c). The resistive micro-heater is 35 μm long, 5 μm wide, and 150nm thick, and is placed parallel and adjacent to the PCW with a separation of 3.25 μm from the heater edge to the PCW center, and is thus far enough to avoid extra optical loss. The current flow results in Joule heating on silicon. The thermal distribution on the device is simulated, as shown in Fig. s3 [26]. The direct contact between the micro-heater and the silicon core layer allows an efficient heat transfer owing to the high thermal conductivity of silicon (k=149 W m$^{-1}$ K$^{-1}$). In addition, the buried oxide layer of SOI wafer functions as a vertical thermal barrier, which further facilitates lateral heat exchange between the heater and the waveguide region [26]. This efficient micro-heater, together with the large TO coefficient of silicon (dn/dT = 1.86×10$^{-4}$ K$^{-1}$ at room temperature), is beneficial for high-speed and low-power operation. Air serves as top cladding, and no passivation layer is required, which simplifies the fabrication steps. The silicon PCW with PC microcavities is fabricated using e-beam lithography and RIE, while gold micro-heater is patterned by photolithography, e-beam evaporation and lift-off process, as illustrated in Fig. s4 [26]. The fabricated device is shown in Figs. 1 (a)-(b) and Fig. s5 (a)-(b) [26].

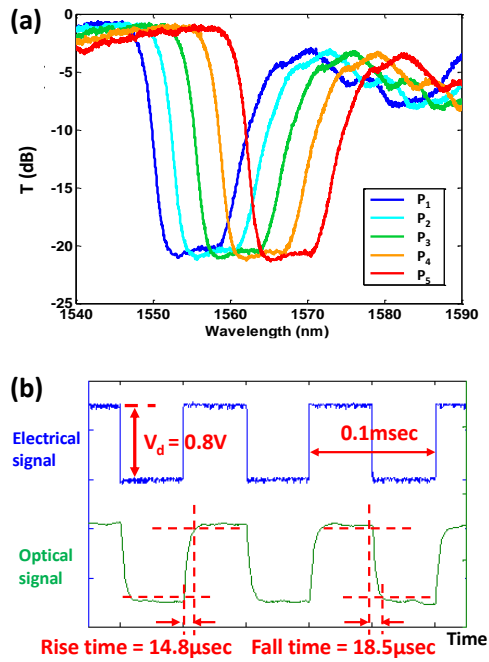

FIG. 4. (a) Measured transmission spectra of the PCW TO switch under different DC driving power, showing a red shift with increased driving power. The measured transmission spectrum is normalized to a reference strip waveguide on the same chip. (b) Measured on-off switching characteristics at 10KHz, with optical wavelength of 1555nm.

A passive optical test is first performed to characterize the coupled PC microcavities. Light from a broadband amplified spontaneous emission (ASE) source is amplified by an Erbium-doped fiber amplifier (EDFA). A polarization controller is used to provide TE-polarized optical input. Light is coupled into and out of the device via grating couplers, and then the optical output is sent to an optical spectrum analyzer. The measured transmission spectrum of the device is normalized to a reference strip waveguide on the same chip and is shown by the dark blue curve in Fig. 4 (a). A resonance with an almost flat bottom is observed from 1552nm to 1558nm. The measured loss on the PCW around 1555nm at the left transmission band is about 1dB.

Next, a DC test is followed to measure the tunability of the TO switch. A constant voltage is applied onto the micro-heater, and the voltage ($V_1$~$V_5$) is tuned from 0V to 0.8V in step of 0.2V, corresponding to the power ($P_1$~$P_5$) of 0mW, 1.2mW, 4.7mW, 10.4mW and 18.2mW (calculated from the measured current and voltage). As a result, the measured transmission spectrum is red-shifted as the power increases, as shown by the overlaid spectra in Fig. 4 (a). This is because the increases of temperature leads to the increase of refractive index of silicon PCW and thus moves the dielectric band farther away from the air band in the photonic band diagram. Alternatively, this can be explained by the increased effective total optical round trip length of the cavity due to the increase of silicon index. Based on Fig. 4 (a), if a laser wavelength is fixed at 1552nm which is the left edge of the flat-bottom resonance, then the required power to switch the optical output power from minimum to maximum is 4.7mW. If a laser wavelength is fixed at the right boundary of the flat-bottom resonance, 1560nm, then the required switching power 18.2mW. This 18.2mW is the constant power for switching from off-state to on-state for all 6nm-wide wavelength range. In the off state, the light is confined in the coupled cavity with the quality (Q) factor of resonator, and finally is released vertically out of plane from the coupled cavity, whereas in the on-state it propagates down the waveguide. The optical extinction ratio measured from Fig. 4 (a) is 20dB. Based on the measured resonance shifts as a function of thermal power applied on the micro-heater, the TO tuning efficiency is about 0.63nm/mW.

Furthermore, to characterize the switching speed and demonstrate the wide operational spectrum range, an on-off switching test is carried out. A tunable laser passing a polarization controller is used to provide TE-polarized optical input, and its wavelength is tuned to 1555 nm which is the center of the flat-bottom resonance in Fig. 4 (a). A square wave RF signal with $V_{pp}$=0.8V at 10KHz is applied on the micro-heater to drive the TO switch. At the optical output end, the switching optical signal is sent to a photodetector and displayed on an oscilloscope, as shown in Fig. 4 (b). The switching rise time (from 10% to 90% transmission) and fall time (from 90% to 10% transmission) are measured to be 14.8 μsec and 18.5 μsec, respectively. Next, the laser wavelength is tuned from 1552nm to 1558nm (from the left edge to right edge of the flat-bottom resonance), and we also observe very similar switching characteristics with almost the same resonance width and extinction ratio, as shown in Fig. s6 [26], verifying the 6nm-wide operational optical bandwidth.

In conclusion, we demonstrate a compact and high-performance TO gate switch based on two coupled L0-type PC microcavities integrated with a micro-heater. The measured transmission spectrum shows a ~6nm-wide flat-bottom resonance with an optical extinction ratio of ~20dB. This flat-bottom resonance is located in a very low group index regime, and thus the optical loss is minimized. This feature is achieved through a very compact PC microcavity (3.78 μm-long). In TO switching experiments, the resonance shifts are measured under different DC driving power with a measured TO tuning efficiency of 0.63nm/mW, and a constant switching power for this 6nm-wide operational spectrum is measured to be 18.2mW. A rise time of 14.8 μsec and a fall time of 18.5 μsec are measured in an on-off switching test. This relatively wide flat-bottom resonance enables significantly broader operational optical bandwidth than widely-used ring resonators and is thus useful for improving traffic management in coarse WDM networks. In the future, to improve the switching speed, the L0-type PC microcavity can be appropriately doped to form a p-n or p-i-n structure and can be operated as a fast electro-optic (EO) switch via free carrier modulation. Since the maxima of the optical field completely overlaps with the central nanohole, one may also consider filling the nanohole with organic EO polymers and applying the device as a hybrid silicon-polymer EO switch or modulator [19]. The coupled PC microcavities can be modified with different resonance dip width (e.g. 1nm, 2nm, etc.) by proper design of microcavity structures and active control of resonant mode-splitting. Furthermore, the present device can be cascaded to enable multiple resonances for more potential applications such as wavelength selective switches and tunable multichannel optical filters.

# APL Supplementary Material

## *Ultra-compact and Wide-spectrum-range Thermo-optic Switch Based on Silicon Coupled Photonic Crystal Microcavities*

Xingyu Zhang,[1,a),b)] Swapnajit Chakravarty,[2,a)] Chi-Jui Chung,[1] Zeyu Pan,[1] Hai Yan,[1] and Ray T. Chen[1,2,a)]

[1]*Department of Electrical and Computer Engineering, University of Texas at Austin, Austin, TX, 78758, USA*
[2]*Omega Optics, Inc., Austin, TX 78757, USA*

**1. Simulation of transmission spectrum as a function of PC microcavity separation**

To investigate the transmission spectrum of the coupled photonic crystal (PC) microcavities when coupled to a W1 photonic crystal waveguide (PCW), successive 3D FDTD simulations are performed for different separation of the studied photonic molecules loaded adjacent to the W1 PCW. Figure S1 shows the evolution of the simulated transmission spectrum as a function the number of PC hole periods (P) between the coupled microcavities. Note that the simulation starts with P=2; P=0 and P=1 cannot exist due to the shifts of $0.2a$=84nm of two adjacent air holes away from each other near each nanohole. At a separation of 9 lattice periods, coupling of the two resonators leads to a 6nm-wide flat-bottom resonance. This coupled L0-type PC microcavities with P=9 provides the widest resonance dip with the shortest cavity length.

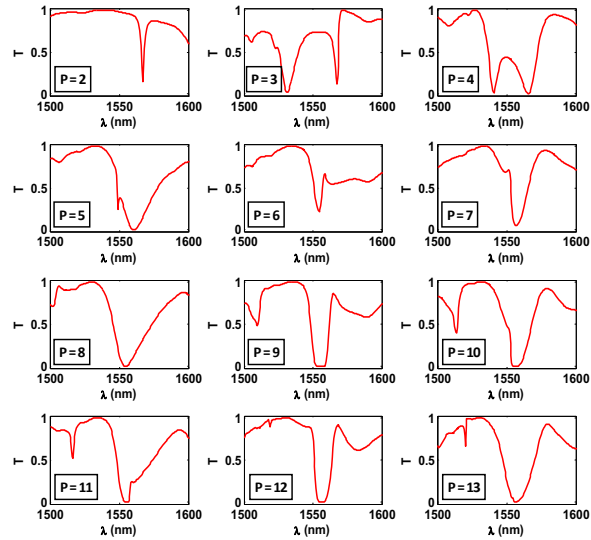

FIG. S1. Evolution of the device transmission spectrum as a function of P, where P represents the number of periods of PC holes between the coupled L0-type PC microcavities.

[a] Authors to whom correspondence should be addressed. Email addresses: xzhang@utexas.edu, swapnajit.chakravarty@omegaoptics.com and chenrt@austin.utexas.edu
[b] Now with Acacia Communications, Inc., Hazlet, New Jersey 07730, USA.

## 2. Simulation of transmission spectrum depending on PC hole diameter

In order to analyze the sensitivity of the device to fabrication imperfections, we investigate the variation of transmission spectrum as a function of the PC hole diameter. Figures S2 (a) and S2 (b) show the simulated transmission spectrum with the hole diameter of d'=191nm and d''=211nm, respectively. Compared with Fig. 2 (b) in the paper, the resonance is shifted by +20nm with the change of diameter by -10nm, while the resonance is shifted by -18nm with the change of diameter by +10nm. The width of resonance dip remains almost unchanged.

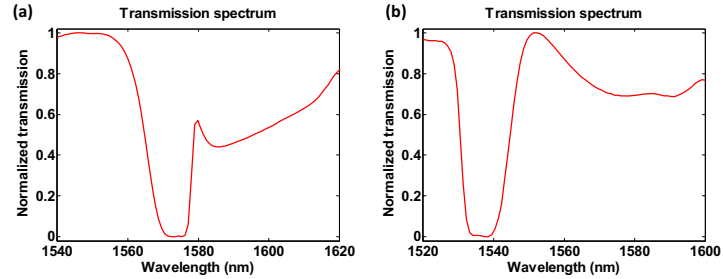

FIG. S2. Simulated transmission spectrum with the PC hole diameter of (a) d'=191nm and (b) d''=211nm, respectively.

## 3. Simulation of thermal distribution around the micro-heater

The schematic of the TO switch is shown in Fig. S3 (a), including gold micro-heater, silicon PCW with coupled PC microcavities, group index tapers and input/output strip waveguides. The thermal distribution across the device induced by the micro-heater is simulated using finite element method. Figure S3 (b) shows the cross-sectional view of the normalized temperature distribution at the center of the device on the SOI substrate.

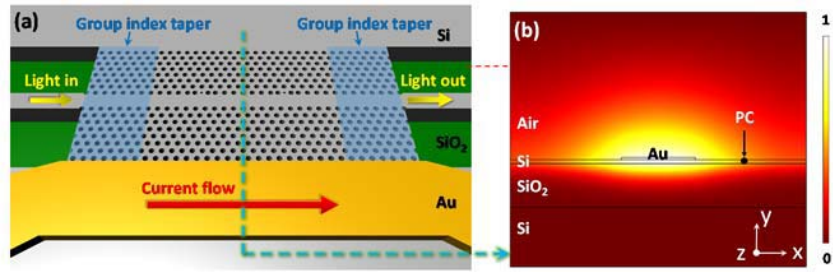

FIG. S3. (a) Schematic of the TO-switch device. (b) Simulated normalized temperature distribution at the cross section of the device (along the dashed lines in (a). Black dot indicates the location of the PCW center.

## 4. Device fabrication

The whole fabrication process is schematically shown in Fig. S4. The fabrication starts with a silicon-on-insulator (SOI) wafer with 250nm-thick silicon top layer and 3μm thick buried oxide [Fig. S4 (a)]. The PCW with cavities is patterned on top silicon layer using e-beam lithography and then through-etched using reactive-ion etching (RIE), as shown in Figs. S4 (b) and S4 (c). Next, a gold micro-heater is patterned by photolithography, e-beam evaporation and lift-off process, as shown in Figs. S4 (d)-S4 (f). A microscope image of a fabricated TO switch is shown in Fig. S5 (a), showing the entire micro-heater. A scanning electron microscope (SEM) image of the entire photonic crystal waveguide is shown in Fig. S5 (b).

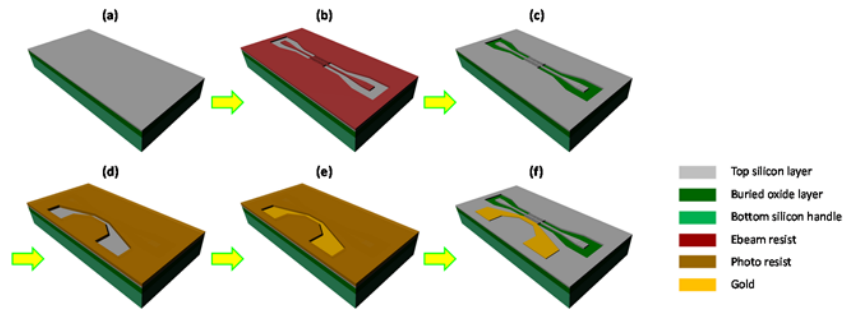

FIG. S4. Fabrication process flow. (a) Silicon-on-insulator wafer. (b) E-beam lithography. (c) RIE. (d) Photolithography alignment. (e) E-beam evaporation. (f) Lift-off.

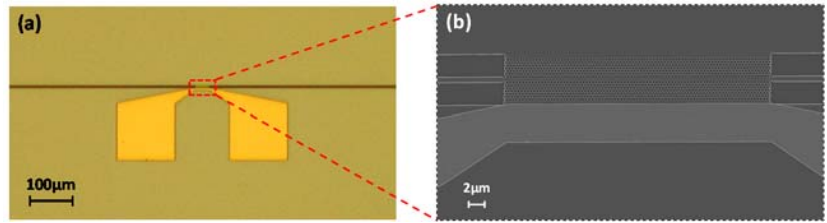

FIG. S5. (a) Microscope images of the fabricated TO switch, showing the entire micro-heater (top view). (b) SEM images of the entire photonic crystal waveguide integrated with the micro-heater.

## 5. Measuring on-off switching characteristics across 6nm spectrum range

Figures. S6 (a) and S6 (b) show the on-off switching characteristics at 1552nm and 1558nm, respectively. When the switch is operated at 1552nm which is the left edge of the flat-bottom resonance, the rise/fall time is slightly decreased, because it requires shorter wavelength shift to switch the optical power between maximum and minimum. In comparison, when the wavelength is farther away from the left edge of the flat-bottom resonance, e.g. 1558nm, the rise/fall time it takes to switch the power on and off is slightly increased. The extinction ratios are almost the same across the 6nm-wide spectrum range, due to the good flatness feature at resonance bottom and almost flat left edge. This relatively wide operational optical bandwidth (6nm) is much better than ring resonator switches whose operation optical bandwidth is usually smaller than 1nm.

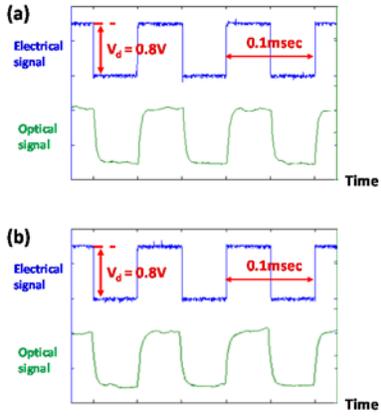

FIG. S6. Measured on-off switching characteristics at RF frequency of 10KHz, with optical wavelength at (a) 1552nm and (b) 1558nm.